\hfuzz 2pt
\vbadness 10000
\font\titlefont=cmbx10 scaled\magstep1

\magnification=\magstep1

\def\asymptotic#1\over#2{\mathrel{\mathop{\kern0pt #1}\limits_{#2}}}

\null
\vskip 1.5cm
\centerline{\titlefont EFFECTIVE DISSIPATIVE DYNAMICS}
\medskip
\centerline{\titlefont FOR POLARIZED PHOTONS}
\vskip 2.5cm
\centerline{\bf F. Benatti}
\smallskip
\centerline{Dipartimento di Fisica Teorica, Universit\`a di Trieste}
\centerline{Strada Costiera 11, 34014 Trieste, Italy}
\centerline{and}
\centerline{Istituto Nazionale di Fisica Nucleare, Sezione di 
Trieste}
\vskip 1cm
\centerline{\bf R. Floreanini}
\smallskip
\centerline{Istituto Nazionale di Fisica Nucleare, Sezione di 
Trieste}
\centerline{Dipartimento di Fisica Teorica, Universit\`a di Trieste}
\centerline{Strada Costiera 11, 34014 Trieste, Italy}
\vskip 2cm
\centerline{\bf Abstract}
\smallskip
\midinsert
\narrower\narrower\noindent
In the framework of open quantum systems, the propagation of polarized
photons can be effectively described using quantum dynamical semigroups.
These extended time-evolutions induce irreversibility and dissipation.
Planned, high sensitive experiments, both in the laboratory
and in space, will be able to put stringent bounds on these 
non-standard effects.
\endinsert

\vfill\eject

The propagation of polarized photons in an optical 
active media is a theoretically
well understood physical process; it can be effectively modelled by
means of linear transformations acting in the space of
polarization states.[1-4] In presence of dissipation however, a more
general formalism in terms of density matrices is usually needed: it can be
physically motivated in the framework of quantum open systems.[5-8]

These systems can be thought as being subsystems in interaction with large
environments. Although the time evolution of the total system follows
the standard quantum mechanical rules, the dynamics of the subsystem,
obtained by the elimination of the environment degrees of freedom,
is usually very involved, developing dissipation and irreversibility.
In many physical instances (essentially when the interaction between
the subsystem and the environment can be considered to be weak)
the subdynamics can be explicitly described in terms of quantum dynamical
semigroups, {\it i.e.} by mean of families of linear maps, transforming
density matrices into density matrices, and satisfying the conditions
of entropy increase, forward in time composition and complete positivity.[5-7]

Various physical situations encountered in quantum optics can be studied
within this framework,[8-10] and indeed, many developments in the theory of
quantum dynamical semigroups have been motivated through these applications.
However, the formalism is very general and has also been applied to describe
many other physical systems, ranging from statistical models,[5-7] 
to molecular systems,[11] to the interaction of a microsystem 
with a measuring apparatus.[12-14]

Further, quantum dynamical semigroups have been recently applied to the
study of dissipation and irreversibility in elementary particle phenomena.[14-16]
The original physical motivation for these investigations come from
quantum gravity: due to the quantum fluctuation 
of the gravitational field and
the presence of virtual black holes, space-time should loose its
``continuum'' aspect at distances of the order of Planck's scale and assume
a ``foam'' like behaviour.[17] As a consequence, new, non-standard phenomena can
arise, leading to loss of quantum coherence.[17-23]

Recent studies based on the dynamics of extended objects (strings and branes)
also support from a more fundamental point of view this possibility.[24, 25]
Unfortunately, our present knowledge of string dynamics does not
allow to quantify precisely the magnitude of the induced non-standard,
dissipative phenomena. In any case, they should produce very small effects;
these are in fact suppressed by at least one inverse power of the Planck mass,
as a rough dimensional analisys suggests. 

Nevertheless, for particular physical systems, involving interference
phenomena, these dissipative effects might be in the reach of present
and future experiments. Indeed, detailed investigations involving
neutral meson systems, neutron interferometry 
and neutrino oscillations
using quantum dynamical semigroups have already allowed to derive
order of magnitude limits on some of the phenomenological constants
parametrizing the new phenomena using current experimental data.[26-29]
More detailed results are expected when new data, in particular involving
correlated neutral mesons, will be available.[30]

Photon interferometry, and more in general optical physics,
is surely another obvious instance in which non-standard,
dissipative effects induced by fundamental dynamics at Planck's scale
might be relevant. Many interferometric-like experiments have been devised
(and will be operational in the near future)
for the study of a wide range of different phenomena, from the analysis
of laser physics, to tests of quantum mechanics, from the detection of
gravitational waves, to the study of astrophysical and cosmological objects;
therefore, it appears relevant to discuss in detail to what extent dissipation
can affect all those observations.

We shall concentrate our attention on the physics of polarized
photons, since it seems to offer many experimental opportunities
for detecting the new, dissipative phenomena.
Quite in general, these effects can be parametrized in terms of six
phenomenological constants, whose presence modify the behaviour of
various physical observables. Explicit expressions for some of these
observables will be given using useful approximations; they can
be of help in fitting experimental data. Indeed, as discussed at the
end, planned, future experimental set-ups, both in the laboratory
and in space, should reach the required sensitivity to measure with
good accuracy at least some of the dissipative parameters.
This is surely an important motivation for further, 
more detailed investigations.

\vskip 1cm

Polarized photons can be effectively described by means of a two-dimensional
Hilbert space, the space of helicity states.[3, 4] Any vector in this space
represents a given polarization and can be identified by two angles
$\theta$ and $\varphi$:
$$
|\theta,\varphi\rangle=\cos\theta\, |+\rangle\,
+\, e^{i\varphi}\sin\theta\, |-\rangle\ ,
\eqno(1)
$$
where $|+\rangle$ and $|-\rangle$ are two orthonormal basis vectors,
representing linearly polarized states. Another convenient basis in
this space is given by the circularly polarized states:
$$
\eqalign{
&|R\rangle={1\over\sqrt2}\Big(|+\rangle\, +\, i |-\rangle\Big)\ ,\cr
&|L\rangle={1\over\sqrt2}\Big(|+\rangle\, -\, i |-\rangle\Big)\ .}
\eqno(2)
$$
With respect to this basis, any (partially) polarized photon state can be
represented by a $2\times2$ density matrix $\rho$; this is a hermitian,
positive operator, {\it i.e.} with positive eigenvalues, and constant trace.
In particular, to the state (1) there corresponds the following matrix:
$$
\rho_{\theta,\varphi}={1\over2}\left[\matrix{
1+\sin\varphi\sin2\theta & \cos2\theta-i\cos\varphi\sin2\theta\cr
\cos2\theta+i\cos\varphi\sin2\theta & 1-\sin\varphi\sin2\theta\cr}\right]\ ;
\eqno(3)
$$
representing a pure state, this matrix is a projector:
$\big(\rho_{\theta,\varphi}\big)^2=\rho_{\theta,\varphi}$.

Quite in general, the dynamics of any state $\rho$ can be described
by an equation of the following form:
$$
{\partial\rho(t)\over \partial t}= -i H\,\rho(t)+i\rho(t)\,
H + {\cal L}[\rho(t)]\ ,
\eqno(4)
$$
where $H$ is the standard hamiltonian, and $\cal L$ is a linear map
that can be written as:
$$
{\cal L}[\rho]=-{1\over2}\sum_j\Big(L^\dagger_jL_j\,\rho +
\rho\, L^\dagger_jL_j\Big)\ 
+\sum_j L_j\,\rho\, L^\dagger_j\ ,
\eqno(5)
$$
while the $L$'s form a suitable collection of operators, such that 
$\sum_j L^\dagger_j L_j$ is a well-defined $2\times 2$ matrix.
Indeed, one can show that this equation is the (unique) 
result of very basic physical
requirements that the complete time-evolution,
$\tau_t:\rho(0)\mapsto \rho(t)$ needs to satisfy; 
the one-parameter (=time) family of
linear maps $\tau_t$ should transform density matrices into density
matrices and have the property of increasing the (von Neumann) entropy,
$S=-{\rm Tr}[\rho(t)\ln\rho(t)]$, of obeying the semigroup composition law,
$\tau_t[\rho(t')]=\rho(t+t')$, for $t,\, t'\geq0$, of being completely
positive. These are the characteristic properties of quantum dynamical 
semigroups, that are therefore generated by equations of the form (4),
with (5).[5-7]

As mentioned in the introductory remarks, quantum dynamical semigroups
have been used to describe a wide range of phenomena related to the
study of open quantum systems; in particular, they have been applied to
analyze non-standard, dissipative effects in the propagation and
decay of neutral meson systems.
Although the basic general motivation behind these treatments is that
quantum phenomena at a fundamental scale produce loss of phase-coherence,
one should always keep in mind that the form (4), (5) of the evolution
equation is very general and independent from the actual microscopic
dynamical mechanism responsible for the dissipative effects:
in view of the properties they satisfy,
any physically sensible description of decoherence phenomena must be based
on equations (4), (5). For these reasons, the discussion in the following,
although applied to a specific model,
retains its validity in a much more general framework.

In connection with these observations, one should add
a further general comment on the time evolution $\rho(t)$. In view of the
interpretation of its eigenvalues as probabilities, the density matrix
$\rho(t)$ needs to be a positive operator for all times; this is clearly
a crucial requirement for the consistency of the whole formalism,
and it is satisfied in all situations only if the map $\rho(0)\mapsto \rho(t)$ 
is completely positive. Roughly speaking, this amounts to the requirement
of positivity for the density matrix of a larger system, involving the
coupling with an extra, auxiliary finite-dimensional system 
(for details, see [5-7, 15, 25]).
This property is trivially satisfied by ordinary (unitary) time evolutions,
and turns out to be crucial in properly treating effects of irreversibility
in correlated systems.[31] For this reason, 
in order to study possible non-standard,
dissipative effect even in simpler, uncorrelated systems
the phenomenological equations (4), (5) should always be used.

In the chosen basis (2), the hamiltonian $H$ has generically the form:
$$
H=\left[\matrix{
\omega_0+\omega_3&\omega_1-i\,\omega_2\cr
\omega_1-i\,\omega_2&\omega_0-\omega_3\cr}\right]\ , 
\eqno(6)
$$
where $\omega_0$ is the average photon energy, while 
the real parameters $\omega_1$, $\omega_2$, $\omega_3$ produce 
the level splitting $\omega\equiv(\omega_1^2+\omega_2^2+\omega_3^2)^{1/2}$.
In the following, we have kept $\omega$ nonvanishing in order to
take into account possible propagation in an optical active
media. Although the most natural way of realizing this is through the
use of a suitable crystal ([3, 4], see also [32, 33]), 
many other unconventional mechanisms leading
to birefringence effects have been discussed in the literature.
They involve the action of external fields,[34] Chern-Simons,[35] 
or more in general Lorentz and $CPT$-violating
modifications of Maxwell Lagrangian,[36]
\phantom{\hskip 2cm}\vfil\eject\noindent
extensions of General Relativity,[37] quantum gravity phenomena.[38]%
\footnote{$^\dagger$}{Other unconventional Planck's scale phenomena
affecting photon propagation, but not directly leading to birefringence, 
have also been studied; see [39] and references therein.}
Furthermore, one should take into account that in general 
the formalism of open quantum systems can also lead to nonvanishing
hamiltonian contributions, so that, even in absence of other
physical mechanisms, birefringence effects should always be present
as the result of the interaction with the environment.[5, 25]

On the other hand, 
the additional piece ${\cal L}[\rho]$ in (5) induces a mixing-enhancing 
mechanism, leading to irreversibility and possible loss of quantum coherence.
Being a linear map, it can be represented as a $4\times4$ matrix
acting on the entries of $\rho$. It can be fully parametrized in terms of
six real phenomenological constants,
$a$, $b$, $c$, $\alpha$, $\beta$, and $\gamma$,
satisfying the following inequalities:
$$
\eqalign{
&2\,R\equiv\alpha+\gamma-a\geq0\ ,\cr
&2\,S\equiv a+\gamma-\alpha\geq0\ ,\cr
&2\,T\equiv a+\alpha-\gamma\geq0\ ,\cr
&RST-2\, bc\beta-R\beta^2-S c^2-T b^2\geq 0\ ,
}\hskip -1cm
\eqalign{
&RS-b^2\geq 0\ ,\cr
&RT-c^2\geq 0\ ,\cr
&ST-\beta^2\geq 0\ ,\cr
&\phantom{\beta^2}\cr
}\eqno(7)
$$
direct consequence of the property of complete positivity.[5, 15]
A convenient explicit expression for the r.h.s. of (4)
can be obtained by decomposing the $2\times2$ density matrix $\rho$
in terms of the Pauli matrices $\sigma_i$, $i=1,2,3$, and the
identity $\sigma_0$:
$$
\rho={1\over2}\sum_{\mu=0}^3 \rho_\mu\, \sigma_\mu\ .
\eqno(8)
$$
One can then rewrite the evolution (4) as a
Schr\"odinger-like equation for the abstract vector
$|\rho(t)\rangle$ of components $(\rho_0,\rho_1,\rho_2,\rho_3)$:
$$
{\partial\over\partial t} |\rho(t)\rangle=-2\, {\cal K}\ |\rho(t)\rangle\ .
\eqno(9)
$$
The $4\times4$ matrix $\cal K$ includes both 
the hamiltonian piece, $-i[H,\rho]$,
and the contribution ${\cal L}[\rho]$, and takes the block-diagonal form:
$$
\big[{\cal K}_{\mu\nu}\big]=\left[\matrix{ 0  & 0\cr
                                           0  & {\cal H}_{ij}\cr}\right]\ ,
\qquad
\big[{\cal H}_{ij}\big]=\left[\matrix{a&b+\omega_3&c-\omega_2\cr
                                     b-\omega_3&\alpha&\beta+\omega_1\cr
                                     c+\omega_2&\beta-\omega_1&\gamma\cr}
									 \right]\ ,
\qquad i,j=1,2,3\ .
\eqno(10)
$$
The formal solution of (9) involves the exponentiation of $\cal K$:
$$
|\rho(t)\rangle= {\cal M}(t)\, |\rho(0)\rangle\ ,\qquad\quad
\big[{\cal M}_{\mu\nu}(t)\big]=\left[\matrix{ 1  & 0\cr
                                           0  & {\cal N}_{ij}(t)\cr}\right]\ ,
\qquad {\cal N}(t)=e^{-2 {\cal H}t}\ .
\eqno(11)
$$
Notice that the time evolution generated by (4), (5) 
is trace-preserving; therefore, 
from the initial normalization condition ${\rm Tr}[\rho(0)]=1$, one immediately
deduces that $\rho_0(t)=1$ for all times, as it is apparent from (11).

Any physical property of a polarized photon beam can be extracted from
the density matrix $\rho(t)$ by computing its trace with suitable
hermitian operators. In particular, the observables that correspond
to the Pauli matrices $\sigma_i$ and the identity $\sigma_0$ give
the so-called (normalized) Stokes polarization parameters.
From the decomposition (8), it is then clear that the vector 
$|\rho\rangle$ precisely represents a normalized Stokes vector;
therefore, the $4\times4$ real matrix ${\cal M}(t)$ in (11) corresponds
to the Mueller matrix connecting the initial Stokes vector $|\rho(0)\rangle$
with the evolved one $|\rho(t)\rangle$ at time $t$.[1-4]

For instance, in absence of the additional piece ${\cal L}[\rho]$,
the matrix ${\cal N}(t)$ in (11) can be explicitly written as:
$$
{\cal N}_{ij}(t)=\delta_{ij}-{\sin2\omega t\over\omega}\, {\cal H}_{ij}
+{2\sin^2\omega t\over\omega^2}\, {\cal H}^2{}_{ij}
\qquad i,j=1,2,3\ .
\eqno(12)
$$
The physical meaning of the corresponding Mueller matrix ${\cal M}(t)$
can be most simply obtained by taking
$\omega_1=\omega_2=\,0$, or alternatively by
switching to the basis in which the hamiltonian $H$ is diagonal. In this case
${\cal M}(t)$ becomes block-diagonal:
$$
{\cal M}(t)=\left[\matrix{1 &&\cr
                             &{\cal R}(t)& \cr
							 && 1\cr}\right]\ ,\qquad
{\cal R}(t)=\left[\matrix{
\cos2\omega t & -\sin2\omega t \cr
\sin2\omega t & \cos2\omega t \cr}\right]\ ;
\eqno(13)
$$
it represents a rotator:
in fact, for linearly polarized states,
$$
|\rho(0)\rangle=\left(\matrix{1 \cr \cos2\theta_0 \cr \sin2\theta_0 \cr 0}
\right)\ ,
\eqno(14)
$$
the direction of polarization, initially along $\theta_0$,
is rotated by an angle $\omega t$, proportional to the elapsed
time.

More in general, any observable $\cal O$ can be decomposed as in (8),
$$
{\cal O}=\sum_{\mu=0}^3 {\cal O}_\mu\, \sigma_\mu\ ,
\eqno(15)
$$
so that its corresponding mean value is given by:
$$
\big\langle{\cal O}(t)\big\rangle\equiv{\rm Tr}[{\cal O}\, \rho(t)]=
\sum_{\mu=0}^3 {\cal O_\mu}\, \rho_\mu(t)\ .
\eqno(16)
$$
Of particular interest is the observable that correspond to the fully
polarized state in (3); the probability that the evolved vector
$|\rho(t)\rangle$ be in such a state is then given by:
$$
{\cal P}_{\theta,\varphi}(t)\equiv
\big\langle\rho_{\theta,\varphi}\big\rangle=
{1\over2}\Big[1+ \rho_1(t)\,\cos2\theta+\rho_2(t)\, \cos\varphi\sin2\theta
+\rho_3(t)\, \sin\varphi\sin2\theta\Big]\ .
\eqno(17)
$$
The corresponding intensity curve that this probability produces can be
compared directly with the experiment, provided explicit expressions
for the entries of the matrix ${\cal M}(t)$ in (11) are given.

Formally, this can be obtained by studying the eigenvalue
problem for the $3\times3$ matrix $\cal H$ in (10):
$$
{\cal H}\, |v^{(k)}\rangle =\lambda^{(k)}\, |v^{(k)}\rangle\ ,
\qquad k=1,2,3\ .
\eqno(18)
$$ 
The three eigenvalues $\lambda^{(1)}$, $\lambda^{(2)}$, $\lambda^{(3)}$
satisfy the cubic equation:
$$
\lambda^3+r\, \lambda^2+ s\, \lambda +w=\,0\ ,
\eqno(19)
$$
with real coefficients,
$$
\eqalignno{
&r\equiv -(\lambda^{(1)}+\lambda^{(2)}+\lambda^{(3)})=
-(a+\alpha+\gamma)\ , &(20a)\cr
&\null\cr
&s\equiv \lambda^{(1)}\lambda^{(2)}+\lambda^{(1)}\lambda^{(3)}+
\lambda^{(2)}\lambda^{(3)}=a\alpha + a\gamma + \alpha\gamma
-b^2-c^2-\beta^2+\omega^2\ , &(20b)\cr
&\null\cr
&w\equiv -\lambda^{(1)}\lambda^{(2)}\lambda^{(3)}=
a(\beta^2-\omega^2_1)+\alpha(c^2-\omega_2^2)+\gamma(b^2-\omega_3^2)\cr
&\hskip 3cm -a\alpha\gamma-2\,bc\beta-2b\omega_1\omega_2-2c\omega_1\omega_3
-2\beta\omega_2\omega_3\ .
&(20c)\cr}
$$
Via Cardano's formula,[40] the corresponding solutions can be expressed in terms 
of the associated discriminant 
${\cal D}=p^3+q^2$, $p=s/3-(r/3)^2$, $q=(r/3)^3-rs/6+w/2$;
the eigenvalues are either all real (${\cal D}\leq 0$), or one
is real and the remaining two are complex conjugate (${\cal D}> 0$).
The degenerate case ${\cal D}=\,0$ occurs when two real eigenvalues
are equal; all three coincide for $p=q=\,0$.

Then, using the fact that the matrix $\cal H$ itself obey the
equation (19), one can show that the entries of $\cal N$ in (11) 
can be written as:
$$
{\cal N}_{ij}(t)=\sum_{k=1}^3 e^{-2\lambda^{(k)} t}
\left[
{\big([\lambda^{(k)}]^2+r\lambda^{(k)}+s\big)\delta_{ij}
+\big(\lambda^{(k)}+r\big){\cal H}_{ij}+{\cal H}^2{}_{ij}\over
3[\lambda^{(k)}]^2+2r\lambda^{(k)}+s}\right]\ ,\quad i,j=1,2,3\ .
\eqno(21)
$$
[The expression in (12) is just a particularly simple example of this
general formula.]

When $\omega=\,0$, {\it i.e.} $\omega_1=\omega_2=\omega_3=\,0$,
the matrix $\cal H$ is real, symmetric and 
non-negative, as guaranteed by the inequalities (7); therefore, 
its eigenvalues are all real and non-negative:
${\cal D}<0$ and $p<0$. Only for sufficiently  large values of
$\omega_1$, $\omega_2$ or $\omega_3$ 
the discriminant $\cal D$ becomes positive and complex
eigenvalues may appear; note that their real part are always non-negative,
since in general the quantum dynamical semigroup generated by
(4), (5) is bounded for any $t$.[41]
Therefore, the general behaviour of the Mueller matrix
${\cal M}(t)$ in (11) depends on the relative magnitude of the constants
$\omega_1$, $\omega_2$ and $\omega_3$ with respect to the dissipative parameters
$a$, $b$, $c$, $\alpha$, $\beta$ and $\gamma$; only when the latter are
small compared to the former an oscillatory behaviour is possible,
while exponential dumping terms prevail, when dissipation is
the dominant phenomena.

In particular, when ${\rm det}({\cal H})\equiv -w\neq0$,
one can show that, in presence of dissipation, 
the real parts of the three eigenvalues
$\lambda^{(1)}$, $\lambda^{(2)}$, $\lambda^{(3)}$ are all 
strictly positive.[29] Then, for large times, the dumping terms dominate
and the Mueller matrix ${\cal M}(t)$ becomes that of a total depolarizer:
$$
{\cal M}(t)
\ \asymptotic\sim\over{t\to\infty}\ {\rm diag}(1,0,0,0)\ .
\eqno(22)
$$
Indeed, in this case the degree of polarization, defined by:
$$
\Pi(t)=\Big(1-{\rm det}[\rho(t)]\Big)^{1/2}\ ,
\eqno(23)
$$
asymptotically vanishes, independently from the initial state $\rho(0)$.

The situation is more complicated in presence of zero eigenvalues, since
the additional cubic condition $w=\,0$ needs to be imposed.
Looking at the explicit expression in $(20c)$ and recalling the
inequalities (7), it is clear that a vanishing $w$ can be obtained
only for very special values of the parameters
$\omega_1$, $\omega_2$, $\omega_3$ and
$a$, $b$, $c$, $\alpha$, $\beta$, $\gamma$.
As a simplifying assumption, let us take $\omega_1=\omega_2=\,0$;
keeping $\omega_3$ arbitrary, the only way to set $w=\,0$ is then to impose
$\gamma=\,0$; indeed, the inequalities (7) further imply 
$b=c=\beta=\,0$ and $a=\alpha$.
Most of the entries of ${\cal N}(t)$ are still exponentially suppressed
for large $t$; however, the presence of the zero eigenvalues now implies
${\cal N}_{33}(t)=1$, so that the asymptotic form of the Mueller matrix
becomes:
$$
{\cal M}(t)
\ \asymptotic\sim\over{t\to\infty}\ {\rm diag}(1,0,0,1)\ .
\eqno(24)
$$
In this case, the degree of polarization (23) vanishes only for states
that initially are linearly polarized, as in (14). 

The large-time behaviours in (22)
and (24) [for the special case $\omega_1=\omega_2=\gamma=\,0$] 
are characteristic of 
the presence of the dissipative contribution (5) to the evolution equation (4)
and can be put to experimental tests. However, for the determination
of the parameters $a$, $b$, $c$, $\alpha$, $\beta$, and $\gamma$,
explicit expressions for the elements of the matrix ${\cal N}(t)$, for
finite $t$, are needed. These are in general very complicated
[{\it cf.} (21)]; nevertheless,
having in mind possible comparisons with experimental data,
their study in suitable approximations seems appropriate.

In order to simplify the treatment, we shall henceforth adopt 
the eigenstates of the hamiltonian in (6) as basis states;
notice that the physical observables, {\it e.g.} the probability in (17),
being the result of a trace operation, are independent from
any choice of basis. [Alternatively, one can assume:
$\omega_1=\omega_2=\,0$.]

When the magnitude of the dissipative, non-standard parameters 
is large or comparable with respect to that of $\omega$, 
a useful working assumption is to take $c$ and $\beta$ 
to be much smaller than the remaining
constants: indeed, this choice is perfectly compatible
with the constraints of complete positivity in (7).
To lowest order, the matrix $\cal H$ becomes block diagonal and a manageable
expression for the entries of the Mueller matrix ${\cal M}(t)$
can be obtained.
Explicitly, ${\cal M}(t)$ can be written as the product:
$$
{\cal M}(t)={\cal M}_D(t)\ \cdot\ {\cal M}_R(t)\ ,
\eqno(25)
$$
where ${\cal M}_D(t)$ is diagonal and contains exponential dumping factors:
$$
{\cal M}_D(t)={\rm diag}\Big(1, e^{-At}, e^{-At}, e^{-2\gamma t}\Big)\ ,
\eqno(26)
$$
while ${\cal M}_R(t)$ is of the form (13), but with the matrix
${\cal R}(t)$ replaced by
$$
{\cal R}_0(t)=\left[\matrix{
\cos2\Omega_0 t+{{\cal R}e(B)\over2\Omega_0}\sin2\Omega_0 t &
-{2\omega+{\cal I}m(B)\over2\Omega_0}\sin2\Omega_0 t\cr
{2\omega-{\cal I}m(B)\over2\Omega_0}\sin2\Omega_0 t &
\cos2\Omega_0 t-{{\cal R}e(B)\over2\Omega_0}\sin2\Omega_0 t}\right]\ ,
\eqno(27)
$$
and
$$
A=\alpha+a\ ,\qquad B\equiv |B| e^{i\phi_B}=\alpha-a+2ib\ ,\qquad
\Omega_0=\sqrt{\omega^2-|B|^2/4}\ .
\eqno(28)
$$
The matrix ${\cal M}_R(t)$ generalizes that of a rotator.
Notice, however, that the oscillator behaviour
depends on the magnitude of $\omega$
with respect to $|B|$; when $\omega<|B|/4$, 
the frequency $\Omega_0$ becomes purely
imaginary and ${\cal M}_R(t)$ contains only exponential terms. 
On the other hand, the form of ${\cal M}_D(t)$ looks like the
Mueller matrix for a random medium;[4]
as in that case, for large times the limit (22) is recovered and any
initial polarization is totally lost.

In practical applications, the initial state $|\rho(0)\rangle$ can often 
be prepared to coincide with that of a linearly polarized photon,
given in (14); one can further set $\theta_0=\,0$, 
by a suitable choice of reference frame.
Then, inserting the previous results into the general expression (17),
the measure of the polarization state along the direction
$\theta$ after a time $t$ would produce
the following intensity pattern:
$$
{\cal P}_\theta(t)={1\over2}\bigg\{1+ e^{-At}\bigg[
\cos2(\theta-\Omega_0 t)+\bigg({|B|\over2\Omega_0}\cos(2\theta+\phi_B)
+\bigg({\omega\over\Omega_0}-1\bigg)\sin2\theta\bigg)
\sin2\Omega_0 t\bigg]\bigg\}\ .
\eqno(29)
$$
As already mentioned, for large times dissipation prevail and all
polarization states become equally probable.

The expression in (29) further simplify when $\gamma=\,0$; as
observed before, this automatically guarantees $c=\beta=\,0$ and further
imposes $b=\,0$ and $a=\alpha$. In this case, one explicitly gets:
$$
{\cal P}_\theta(t)={1\over2}\Big\{1+ e^{-2\alpha t}
\cos2(\theta-\omega t)\Big\}\ .
\eqno(30)
$$
This is the most simple expression that the transition probability
${\cal P}_\theta(t)$ takes in presence of dissipative effects.

Another useful approximation of the general formula (17) can be obtained 
when the non-standard parameters $a$, $b$, $c$, $\alpha$, $\beta$, and $\gamma$
are small compared with $\omega$, assumed to be non-vanishing.
In this case, the additional piece
${\cal L}[\rho]$ in the evolution equation (4) can be treated as
a perturbation.[15] All entries in the $3\times3$ matrix ${\cal N}(t)$
are now non-vanishing and in general, the resulting Mueller matrix
in (11) can not be decomposed as a product of simpler matrices, as in (25).
Explicit expressions for its entries, expanded up to second
order in the small parameters, are collected in the Appendix.
Using these results, the transition probability ${\cal P}_\theta(t)$
takes the form:
$$
\eqalign{
{\cal P}_\theta(t)={1\over2}\bigg\{1+ e^{-At}\bigg[
\cos2(\theta &-\Omega t)+\bigg({|B|\over2\Omega}\cos(2\theta+\phi_B)
+\bigg({\omega\over\Omega}-1\bigg)\sin2\theta\bigg)\sin2\Omega t\cr
&+{|C|^2\over\Omega^2}\sin\phi_C\bigg(\sin(2\theta+\phi_C)\cos2\Omega t
-\sin\phi_C\cos2\theta\bigg)
\bigg]\bigg\}\ ,}
\eqno(31)
$$
where $A$ and $B$ are as in (28), while
$$
C\equiv|C|e^{i\phi_C}=c+i\beta\ ,\qquad \Omega=\sqrt{\omega^2-|C|^2-|B|^2/4}\ .
\eqno(32)
$$
In writing (31), we have reconstructed the exponential
factors by consistently putting together the terms linear and 
quadratic in $t$; a similar treatment has allowed writing the oscillatory
contributions in terms of the frequency $\Omega$.

It is worth noting that the expression (31) reduces to the in (29)
for $|C|=\,0$, {\it i.e.} when $c=\beta=\,0$: 
it is therefore a correction to (29) for nonvanishing $C$.%
\footnote{$^\dagger$}{Analogously, in the same limit
the elements of ${\cal M}(t)$ listed in the Appendix reduces
to those in $(25)-(27)$.}
In this respect, the validity of (31) goes beyond the approximation
in which it has been derived: it can be considered as the expansion
of the full probability ${\cal P}_\theta(t)$ up to second order 
in $c$ and $\beta$, and thus it is valid also for vanishing $\omega$. 

The behaviour of the probabilities in $(29)-(31)$, and of other observables
that can be similarly constructed, are clearly affected by the presence 
of dissipation and irreversibility.
From the experimental point of view, the actual visibility of such 
non-standard effects clearly depends on the magnitude
of the parameters $a$, $b$, $c$, $\alpha$, $\beta$, and $\gamma$.
In a phenomenological approach, it is hard to give an apriori
estimate on how large the dissipative effects should be.
However, as already mentioned in the introductory remarks,
a general framework in which dissipation naturally emerges is provided
by the study of subsystems in interaction with large environments.
In such instances, the non-standard effects can be roughly estimated
to be proportional to the typical energy of the system,
while suppressed by inverse powers of the characteristic energy
scale of the environment.

In the case of polarized photon beams, these considerations,
together with the general idea that dissipation is induced by
quantum effects at Planck's scale, lead to predict very small values
for the parameters $a$, $b$, $c$, $\alpha$, $\beta$, and $\gamma$;
for any fixed observational condition, an upper bound on the magnitude 
of these parameters can be roughly evaluated to be of order
$E^2/M_P$, with $E$ the average photon energy and $M_P$ the Planck mass.
This ratio is of order $10^{-49}\ {\rm GeV}$ for a
typical radio-wave, $10^{-38}\ {\rm GeV}$ for ordinary laboratory
laser beams, and $10^{-19}\ {\rm GeV}$ or more for energetic
$\gamma$-rays.

At first sight, it might look very hard to construct an actual experimental
set-up sensible to such tiny values. However, the sophistication of
present and planned ``optical'' devices is so high [42] that at least some bounds
on $a$, $b$, $c$, $\alpha$, $\beta$, and $\gamma$ should actually be
obtainable in the near future. In fact, at least in principle,
a very simple set-up is needed in order to test the
non-standard dynamics in (4), (5). An initially polarized photon beam 
evolves undisturbed for a time $t$; 
its final polarization state is then determined
by measuring the corresponding Stokes parameters, and compared with
the results in (11). In practice, it might be more convenient to perform
an interferometric polarization test; the resulting intensity
curve can be directly compared with the behaviour of ${\cal P}_\theta(t)$
in $(29)-(31)$.

As a simplifying working assumption, 
take $\gamma=\,0$, 
so that only the parameter $\alpha$ survives, as a consequence of (7);
then, from the expression in (30), the ability of detecting
the non-standard effects is connected to the sensitivity
in isolating the exponential factor 
$e^{-2\alpha t}$ from the experimental data.%
\footnote{$^\dagger$}{As already mentioned, also $\omega$
contains in general dissipative contributions, resulting from the
interaction with the environment. However, these contributions to
birefringence can not be disentangle from those produced by 
other physical effects (see [32-38]).
On the other hand, the dependence of ${\cal P}_\theta$
on the non-standard parameters $a$, $b$, $c$, $\alpha$, $\beta$, and $\gamma$
is distinctive of dissipative phenomena and can not be mimicked by
other unconventional mechanisms.}
This can be very high, so that, for long
enough $t$, a good sensitivity on $\alpha$ can be reached.

Although a detailed analysis of possible experimental
set-ups that can be used for the measure of $\alpha$
is surely beyond the scope of the present investigation,
some general considerations on the sensitivity of present and future
apparatus can be given.
In the case of ground-based laboratory experiments using ordinary laser
beams, even for relatively large ``storage'' times $t$,
as obtained in high efficient cavities [42, 43]
or in the interferometric detectors for gravitational waves,[44]
the sensitivity
on $\alpha$ can not optimistically exceed $10^{-30}\ {\rm GeV}$,
a few orders of magnitude away from the estimated upper bound on $\alpha$.

The situation clearly improves using high energetic polarized
$\gamma$-beams (with $E\sim 10^2\ \rm {GeV}$ or greater), 
as the ones expected in the so-called Photon Colliders,[45] since now
one should have $\alpha\leq 10^{-15}\ {\rm GeV}$; the required sensitivity
for measuring this parameter from (29) can be reached with a
path-length $\ell\equiv t$ of just a few centimeters.
These high-energy accelerators, together with the $e^+-e^-$ linear colliders
to which they are coupled, turn out to be particularly suited for studying
certain aspects of the Standard Model;[46] many projects for their actual
realization are in advanced stage of development.[47]
On the other hand, the construction of polarized photon beams of more modest 
energies ($E\sim 1\ \rm {GeV}$)
is surely in the capacity of any high energy laboratory
({\it e.g}, see the discussion presented in [48]).
They might actually soon be built with the aim of studying the 
birefringence effects predicted in [32]; clearly, they would also 
provide a suitable venue for deriving
accurate bounds on some of the parameters
$a$, $b$, $c$, $\alpha$, $\beta$, and $\gamma$.

Polarized photons of astrophysical and cosmological origin
can also be used to probe the presence of the dissipative
effects described before. Indeed, radio-signals from
active galactic nuclei and quasars have already been used to put
stringent bounds on $CPT$-violating birefringence effects.[35]
For a typical radio-wave (with frequency $\sim1\ {\rm GHz}$), 
the dimensional arguments
discussed before would give un upper bound on the magnitude of the
dissipative effects that is really very small: 
$\alpha\leq 10^{-49}\ {\rm GeV}$; however, the propagation time $t$ 
can now be as big as the inverse of the Hubble constant.
Therefore, the sensitivity of the present radio-telescope polarization
measures is not too far form the above upper limit 
and improvements can be expected in the future.[49]

The development of very sophisticated detectors, both ground-based
and space-based, has allowed the observation and the study of
$\gamma$-ray emissions from a variety of astrophysical sources.[50, 51]
In view of their extremely high energy 
(ranging from $10^2\ {\rm KeV}$ up to $1\ {\rm TeV}$ or more) 
and of their extragalactic or
cosmological origin (resulting in large
propagation times $t$), these photons turn out to be a very interesting
system for measuring non-standard, dissipative effects.
In fact, the physical mechanisms that have been proposed to explain
the origin of these energetic emissions give rise to (partially)
polarized photons,[50-52] and preliminary observations seem to confirm
this prediction.[53, 54]
If efficient polarimeters will be coupled to the next generation
of orbiting $\gamma$-ray spectrometers, accurate measurements of the
parameters in (10) might indeed be possible.

In conclusion,
the study of polarized photon beams can provide very useful information
on the presence of dissipation and irreversibility induced by
a fundamental ``stringy'' dynamics. Future experiments, in the laboratory
and in space, will likely be able to put stringent bounds on
these non-standard effects.

\vskip 1cm

\line{\bf Appendix\hfill}
\bigskip

As discussed in the text, when the parameters
$a$, $b$, $c$, $\alpha$, $\beta$, and $\gamma$ can be considered small
with respect to $\omega$, perturbation theory can be used to find 
a convenient explicit expression for the solution of the equation (9). 
Up to second order in the small
parameters, the entries of the matrix ${\cal N}(t)$ in (11) can then
be written as:
$$
\eqalign{
&{\cal N}_{11}(t)=e^{-At}\Bigg[\cos2\Omega t
+{{\cal R}e(B)\over2\Omega}\sin2\Omega t\Bigg]
-2\Bigg[{{\cal I}m(C)\over\Omega}\Bigg]^2\sin^2\Omega t\ ,\cr
&{\cal N}_{12}(t)=-e^{-At}\Bigg[{2\omega+{\cal I}m(B)\over2\Omega}\Bigg]
\sin2\Omega t
+{{\cal I}m(C^2)\over2\Omega^2}\cos2\Omega t\ ,\cr
&{\cal N}_{13}(t)=
{{\cal I}m(C)\over\Omega} e^{-2\gamma t}
-e^{-At}{|C|\over\Omega}\sin(2\Omega t+\phi_C)\cr
&\hskip 3cm
-\bigg[|C|(A-2\gamma)\sin(\Omega t+\phi_C)+{\cal R}e(BC)\sin\Omega t\bigg]
{\sin\Omega t\over\Omega^2}\cr
&{\cal N}_{22}(t)=e^{-At}\Bigg[\cos2\Omega t
-{{\cal R}e(B)\over2\Omega}\sin2\Omega t\Bigg]
-2\Bigg[{{\cal R}e(C)\over\Omega}\Bigg]^2\sin^2\Omega t\ ,\cr
&{\cal N}_{23}(t)=
-{{\cal R}e(C)\over\Omega} e^{-2\gamma t}
+e^{-At}{|C|\over\Omega}\cos(2\Omega t+\phi_C)\cr
&\hskip 3cm
+\bigg[|C|(A-2\gamma)\cos(\Omega t+\phi_C)+{\cal I}m(BC)\sin\Omega t\bigg]
{\sin\Omega t\over\Omega^2}\ ,\cr
&{\cal N}_{33}(t)=e^{-2\gamma t} + {2|C|^2\over\Omega^2}
\sin^2\Omega t\ ,\cr
}
$$
where the definition (28) and (32) have been used; the remaining off-diagonal
elements ${\cal N}_{21}$, ${\cal N}_{31}$ and ${\cal N}_{32}$
can be obtained from ${\cal N}_{12}$, ${\cal N}_{13}$ and ${\cal N}_{23}$,
respectively, by letting $\omega\rightarrow-\omega$ and 
$\Omega\rightarrow-\Omega$.

\vfill\eject

\centerline{\bf REFERENCES}
\bigskip\medskip

\item{1.} M. Born and E. Wolf, {\it Principles of Optics},
(Pergamon Press, Oxford, 1980)
\smallskip
\item{2.} L.D. Landau and E.M. Lifshitz, {\it The Classical Theory of Fields},
(Pergamon Press, New York, 1975); {\it Quantum Mechanics},
(Pergamon Press, New York, 1975)
\smallskip
\item{3.} E. Collett, {\it Polarized Light}, (Marcel Dekker, New York, 1993)
\smallskip
\item{4.} C. Brosseau, {\it Fundamentals of Polarized Light},
(Wiley, New York, 1998)
\smallskip
\item{5.} R. Alicki and K. Lendi, {\it Quantum Dynamical Semigroups and 
Applications}, Lect. Notes Phys. {\bf 286}, (Springer-Verlag, Berlin, 1987)
\smallskip
\item{6.} V. Gorini, A. Frigerio, M. Verri, A. Kossakowski and
E.C.G. Surdarshan, Rep. Math. Phys. {\bf 13} (1978) 149 
\smallskip
\item{7.} H. Spohn, Rev. Mod. Phys. {\bf 53} (1980) 569
\smallskip
\item{8.} W.H. Louisell, {\it Quantum Statistical Properties of Radiation},
(Wiley, New York, 1973)
\smallskip
\item{9.} C.W. Gardiner, {\it Quantum Noise} (Springer, Berlin, 1992)
\smallskip
\item{10.} M.O. Scully and M.S. Zubairy, 
{\it Quantum Optics} (Cambridge University Press, Cambridge, 1997)
\smallskip
\item{11.} D.K. Kilin, The role of the environment in molecular systems,
{\tt quant-ph/0001004}
\smallskip
\item{12.} L. Fonda, G.C. Ghirardi and A. Rimini, Rep. Prog. Phys.
{\bf 41} (1978) 587 
\smallskip
\item{13.} H. Nakazato, M. Namiki and S. Pascazio,
Int. J. Mod. Phys. {\bf B10} (1996) 247
\smallskip
\item{14.} F. Benatti and R. Floreanini, Phys. Lett. {\bf B428} (1998) 149
\smallskip
\item{15.} F. Benatti and R. Floreanini, Nucl. Phys. {\bf B488} (1997) 335
\smallskip
\item{16.} F. Benatti and R. Floreanini, Nucl. Phys. {\bf B511} (1998) 550
\smallskip
\item{17.} S. Hawking, Comm. Math. Phys. {\bf 87} (1983) 395; Phys. Rev. D
{\bf 37} (1988) 904; Phys. Rev. D {\bf 53} (1996) 3099;
S. Hawking and C. Hunter, Phys. Rev. D {\bf 59} (1999) 044025
\smallskip
\item{18.} J. Ellis, J.S. Hagelin, D.V. Nanopoulos and M. Srednicki,
Nucl. Phys. {\bf B241} (1984) 381; 
\smallskip
\item{19.} S. Coleman, Nucl. Phys. {\bf B307} (1988) 867
\smallskip
\item{20.} S.B. Giddings and A. Strominger, Nucl. Phys. {\bf B307} (1988) 854
\smallskip
\item{21.} M. Srednicki, Nucl. Phys. {\bf B410} (1993) 143
\smallskip
\item{22.} W.G. Unruh and R.M. Wald, Phys. Rev. D {\bf 52} (1995) 2176
\smallskip
\item{23.} L.J. Garay, Phys. Rev. Lett. {\bf 80} (1998) 2508;
Phys. Rev. D {\bf 58} (1998) 124015
\smallskip
\item{24.} J. Ellis, N.E. Mavromatos and D.V. Nanopoulos, Phys. Lett.
{\bf B293} (1992) 37; Int. J. Mod. Phys. {\bf A11} (1996) 1489
\smallskip
\item{25.} F. Benatti and R. Floreanini, Ann. of Phys. {\bf 273} (1999) 58
\smallskip
\item{26.} F. Benatti and R. Floreanini, Phys. Lett. {\bf B401} (1997) 337
\smallskip
\item{27.} F. Benatti and R. Floreanini, Phys. Lett. {\bf B465}
(1999) 260
\smallskip
\item{28.} F. Benatti and R. Floreanini, Phys. Lett. {\bf B451} (1999) 422
\smallskip
\item{29.} F. Benatti and R. Floreanini, JHEP {\bf 02} (2000) 032
\smallskip
\item{30.} F. Benatti and R. Floreanini, $CPT$, dissipation, and all that,
in {\it Physics and Detectors for Daphne}, ``Frascati Physics Series'', vol. XVI,
1999, p. 307, {\tt hep-ph/9912426}
\smallskip
\item{31.} F. Benatti and R. Floreanini,
Mod. Phys. Lett. {\bf A12} (1997) 1465; 
Banach Center Publications, {\bf 43} (1998) 71; 
Phys. Lett. {\bf B468} (1999) 287; On the weak-coupling limit and complete
positivity, Chaos, Sol. Fractals, to appear 
\smallskip
\item{32.} N. Cabibbo, G. Da Prato, G. De Franceschi and U. Mosco,
Phys. Rev. Lett. {\bf 9} (1962) 435
\smallskip
\item{33.} V.A. Maisheev, Sensitivity of single crystals to the
circular polarization of high-energy $\gamma$-rays, {\tt hep-ph/9912437}
\smallskip
\item{34.} V. G. Baryshevsky, JHEP {\bf 04} (1998) 018;
Phys. Lett. {\bf A260} (1999) 24
\smallskip
\item{35.} S. Carroll, G. Field and R. Jackiw, Phys. Rev. D {\bf 41}
(1990) 1231
\smallskip
\item{36.} D. Colladay and V.A. Kostelecky, Phys. Rev. D {\bf 58} (1998) 116002
\smallskip
\item{37.} M.D. Gabriel, M.P. Haugan, R.B. Mann and J.H. 
Palmer, Phys. Rev. D {\bf 43}(1991) 308; 
S.K. Solanki, M.P. Haugan and R.B. Mann,
Phys. Rev. D {\bf 59} (1999) 047101;
M.P. Haugan and T.F. Kauffmann, Phys. Rev. D {\bf 52}(1995) 3168;
R.-G. Cai, Nucl. Phys. {\bf B524} (1998) 639;
N.L. Lepora, Cosmological birefringence and the microwave background,
{\tt gr-qc/9812077}
\smallskip
\item{38.} R. Gambini and J. Pullin, Phys. Rev. D {\bf 59} (1999) 124021
\smallskip
\item{39.} G. Amelino-Camelia, Gravity-wave interferometers as 
probes of a low-energy effective quantum gravity,
{\tt gr-qc/9903080}
\smallskip
\item{40.} For instance, see: M. Artin, {\it Algebra}, (Prentice Hall,
Englewood Cliffs (NJ), 1991)
\smallskip
\item{41.} K. Lendi, J. Phys. {\bf A 20} (1987) 13
\smallskip
\item{42.} H.-A. Bachor, {\it A Guide to Experiments in Quantum Optics},
(Wiley-VCH, Weinhein, 1998)
\smallskip
\item{43.} R. Cameron {\it et al.}, Phys. Rev. D {\bf 47} (1993) 3707;
A.M. De Riva {\it et al.}, Rev. Sci. Instrum. {\bf 67} (1996) 2680;
D. Bakalov {\it et al.}, Quantum Semiclass. Opt. {\bf 10} (1998) 239
\smallskip
\item{44.} P.R. Saulson, {\it Fundamentals of Interferometric Gravitational 
Wave Detectors}, (World Scientific, Singapore, 1994)
\smallskip
\item{45.} V. Telnov, High energy photon colliders, {\tt hep-ex/0001029}
\smallskip
\item{46.} E. Accomando et al. (ECFA/DESY LC Physics Working Group), 
Phys. Rept. {\bf 299} (1998) 1
\smallskip
\item{47.} The NLC Design Group, Zeroth-Order Design Report 
for the Next Linear Collider, SLAC Report 474, 1996;\hfill\break
R. Brinkmann {\it et al.}, Nucl. Instr. Meth.
{\bf A406} (1998) 13;\hfill\break
N. Akasaka {\it et al.}, JLC Design Study, KEK-REP-97-1, 1997
\smallskip
\item{48.} G. Cantatore, F. Della Valle, E. Milotti, L. Dabrowski and C. Rizzo,
Phys. Lett. {\bf B265} (1991) 418
\smallskip
\item{49.} K. Rohlfs and T.L. Wilson, {\it Tools of Radio Astronomy},
(Springer Verlag, New York, 1999)
\smallskip
\item{50.} C.M. Hoffman, C. Sinnis, P. Fleury and M. Punch,
Rev. Mod. Phys. {\bf 71} (1999) 897
\smallskip
\item{51.} T. Piran, Phys. Rep. {\bf 314} (1999) 575
\smallskip
\item{52.} A. Gruzinov and E. Waxman, Astroph. J. {\bf 511} (1999) 852, 
{\tt astro-ph/9807111};
A. Gruzinov, Astroph. J. {\bf 525} (1999) L29, {\tt astro-ph/9905276}
\smallskip
\item{53.} S. Covino {\it et al.}, Astron. Astroph. {\bf 348} (1999) L1,
{\tt astro-ph/9906319}
\smallskip
\item{54.} R.A.M.J. Wijers {\it et al.}, Astroph. J. {\bf 523} (1999) L33,
{\tt astro-ph/9906346}

\bye